\documentclass[epjST]{svjour}
\usepackage{graphics}
\usepackage{xcolor}
\begin{document}
\title{Chiral phase transition and kaon-to-pion ratios in the entanglement SU(3) PNJL model}
\author{Blaschke D.\inst{1, 2, 3}\fnmsep\thanks{\email{david.blaschke@gmail.com}} , Friesen A. V.\inst{1}\fnmsep\thanks{\email{avfriesen@theor.jinr.ru}} \and Kalinovsky Yu.\,L.\inst{1, 4}\fnmsep\thanks{\email{kalinov@jinr.ru}} \and Radzhabov A.\inst{5}\fnmsep\thanks{\email{aradzh@icc.ru}}}
\institute{Joint Institute for Nuclear Research, 141980, Dubna, Russia \and Institute for Theoretical Physics, University of Wroc{\l}aw, 50-204 Wroc{\l}aw, Poland \and National Research Nuclear University (MEPhI), 115409 Moscow, Russia \and Dubna State University, 141982, Dubna, Russia \and 
Matrosov Institute for System Dynamics and Control Theory, Irkutsk 664033, Russia}
\abstract{
Within the three-flavor PNJL and EPNJL chiral quark models we have obtained pseudoscalar meson properties in quark matter at finite temperature $T$ and baryochemical potential $\mu_B$. 
We compare the meson pole (Breit-Wigner) approximation with the Beth-Uhlenbeck (BU) approach that takes into account the continuum of quark-antiquark 
scattering states when determining the partial densities of pions and kaons.
We evaluate the kaon-to-pion ratios along the (pseudo-)critical line in the $T-\mu_B$ plane as a proxy for the chemical freezeout line, whereby the variable 
$x=T/\mu_B$ is introduced that corresponds to the conserved entropy per baryon as initial condition for the heavy-ion collision experiments.
We present a comparison with the experimental pattern of kaon-to-pion ratios within the BU approach and using $x$-dependent pion and strange quark potentials.
A sharp "horn" effect in the energy dependence $K^+/\pi^+$ ratio is explained by the enhanced pion production at energies above $\sqrt{s_{NN}}=8$ GeV, when
the system enters the regime of meson dominance.
This effect is in line with the enhancement of low-momentum pion spectra that is discussed as a precursor of the pion Bose condensation and entails the 
occurrence of a nonequilibrium pion chemical potential of the order of the pion mass.
We elucidate that the horn effect is not related to the existence of a critical endpoint in the QCD phase diagram.
} 
\maketitle

\section{Introduction}
\label{intro}
Quantum chromodynamics predicts the existence of a quark-gluon plasma (QGP) phase, where hadronic bound states dissociate as a consequence of the restoration of chiral symmetry which melts the chiral condensate and the quark mass gap so that a system of deconfined quarks and gluons emerges. 
The study of this phase transition in ab-initio simulations of lattice QCD (LQCD) thermodynamics is yet limited to finite temperatures and low chemical potentials only \cite{Bazavov:2018mes,Borsanyi:2020fev}, so that the development of effective model descriptions is of importance.

In the present work, we discuss the chiral phase transition and the in-medium behaviour of pseudoscalar mesons in the framework of the SU(3) NJL model extended by the coupling to the Polyakov loop, the PNJL model. 
This model is capable of describing both the chiral and the deconfinement transition \cite{Ratti:2005jh,Fukushima:2008wg}, as well as the low-lying hadron spectrum \cite{Blanquier:2011zz}. 
By the mechanism of dynamical chiral symmetry breaking a  chiral condensate develops and quarks acquire large quasiparticle masses. 
The chiral symmetry gets restored when the dynamically generated quark masses drop as a function of temperature and chemical potentials. 
Within this framework the confinement of coloured quark states is effectively taken into account by coupling the chiral quark dynamics to the Polyakov loop and its effective potential.

In the PNJL model, the correlation between quark and gauge fields appears only by the minimal coupling through the covariant derivative 
but is nevertheless strong enough to synchronize the two transitions \cite{Ratti:2005jh}.
However, the absolute value of the pseudocritical temperature of the chiral crossover transition with above 200 MeV is too large when compared 
to the lattice QCD result for 2+1 flavors of $T_c=156.5\pm 1.5$ MeV \cite{Bazavov:2018mes}.
Different solutions of this problem are suggested in the literature.
Nonlocal chiral quark models with covariant formfactors fitted to QCD running masses and wave function renormalization have much smaller pseudocritical temperatures of $\sim 110-140$ MeV for the chiral transition \cite{Blaschke:2000gd,GomezDumm:2005hy,Contrera:2009hk,Contrera:2016rqj}. 
When their chiral quark dynamics gets coupled to the Polyakov loop a strong improvement is obtained 
\cite{Contrera:2016rqj,Contrera:2007wu,Radzhabov:2010dd}.
Unfortunately, the transition becomes strong and even of first order with a diverging chiral susceptibility \cite{Horvatic:2010md}.

Another, even more physical solution is to go beyond the mean field approximation and to consider the role of hadronic excitations in the medium in 
melting the chiral condensate  \cite{Jankowski:2012ms}.
This leads to an excellent agreement with lattice QCD up to and including the pseudocritical temperature, but there are two problems which have not yet been consistently solved: 
the hadrons should dissociate into their quark and gluon constituents at the chiral/deconfinement transition and simultaneously these degrees of freedom shall appear as quasiparticles in the system without generating discontinuities in the behaviour of the order parameters and their derivatives.
This is a formidable task  that requires a selfconsistent formulation and subsequent solution. First steps in this direction have been explored, e.g., within a generalized Beth-Uhlenbeck approach \cite{Blaschke:2016fdh,Blaschke:2017pvh,Bastian:2018wfl}.

An intermediate solution is the introduction of an improved mean field description of the chiral transition.
It is the so-called entanglement PNJL (EPNJL) model, where the scalar four-quark interaction $g_S$ 
is modified by an additional dependence on the Polyakov loop \cite{Sakai:2010rp,Ruivo:2012xt}. 
Such entanglement leads to a change of the phase diagram: the (pseudo-)critical temperature of the chiral and deconfinement crossover transition at low chemical potentials is lowered towards the value of the Lattice QCD prediction. 
Because of a relationship between the Polyakov loop expectation value in the confined phase and the hadronic spectrum \cite{Megias:2012kb} 
one may regard this kind of model as mimicking the backreaction of hadronic modes in the medium on the quark thermodynamics.

Including the vector meson interaction in the model leads to a dependence of the equation of state on the vector coupling constant $g_V$. 
When the vector interaction is taken into account, the critical endpoint (CEP) of first order transitions appears at a higher chemical potential and lower temperature \cite{Stephanov:2007fk} or even vanishes completely from the phase diagram \cite{Sasaki:2006ws,Bratovic:2012qs,Friesen:2011wt,Friesen:2014mha,Chu:2016ixv}.

The study of pseudo-scalar mesons is interesting since their Goldstone boson nature is associated with the breaking of the chiral symmetry and therefore sensible to its restoration in the medium. 
One of the central problems in the investigation of the transition from hadronic to quark matter is a description of the dissociation of hadrons into their quark constituents. 
The PNJL model is conventionally used in the mean field approximation to address the thermodynamic behaviour of the chiral condensate and the Polyakov loop as order parameters of chiral symmetry breaking and color SU(3) center symmetry breaking, respectively.
Including Gaussian fluctuations, it is able to describe the transition of pseudoscalar meson bound states to the quark matter continuum, but the quark-meson 
\cite{Friesen:2013bta} and meson-meson \cite{Quack:1994vc} 
correlations, which can play an important role in the quark-hadron transition appear only at higher orders. 

The relativistic Beth-Uhlenbeck (BU) approach to the description of mesonic bound and scattering states in a quark plasma has been developed in 
\cite{Hufner:1994ma,Wergieluk:2012gd,Blaschke:2013zaa} for the two-flavor case.
Within the SU(3) PNJL model we have developed such a BU approach to scalar and pseudoscalar mesons in \cite{Dubinin:2016wvt}
and further extend it here by taking also into account the Kobayashi-Maskawa-'t Hooft determinant interaction term. 

In the BU approach 
one obtains not only the mass splitting of the charge multiplets in hot and dense matter that is also seen in the BW approximation to solutions of the in-medium Bethe-Salpeter equation \cite{Costa:2005cz,Blaschke:2011yv,Friesen:2018ojv}. 
In this approach it is clearly seen from the behaviour of the corresponding phase shifts for pseudoscalar mesons as functions of energy that 
there are scattering state contributions from the quark-antiquark continuum and 
for finite temperatures and baryon chemical potentials an unusual plasmon mode appears in the $K^+$ channel which develops a bound state pole under conditions that would correspond to the chemical freeze-out in the region of the "horn" \cite{Dubinin:2016wvt}. 
As we have shown in Ref.~\cite{Friesen:2018ojv}, a decisive role in the description of the different patterns of the $K^+/\pi^+$ and $K^-/\pi^-$ ratios as a 
function of $\sqrt{s_{NN}}$ is played by the chemical potentials that are conjugate to (quasi) conserved charges in the system.

The energy scan of the kaon-to-pion ratios has a different pattern depending on the electric charge of the mesons. 
For the ratio of positively charged strange to nonstrange mesons a ''horn'' was found at the collision energy $\sqrt{s_{NN}}\sim$ 7-10 GeV
by the NA49 experiment at CERN SPS in an energy scan with fixed-target Pb+Pb collisions.
The ratio of negatively charged  mesons shows a monotonously rising behaviour.
See Fig. 1 of \cite{Aduszkiewicz:2019zsv} for the world data on the energy dependence of  the kaon-to-pion ratios.
The ''horn'' is supposed to be a signal of the formation of the QGP phase during the heavy-ion collision \cite{Afanasiev:2002mx,Alt:2007aa,Adamczyk:2017iwn,Andronic:2008gu}. 
The original prediction of the effect by Gazdzicki and Gorenstein 
\cite{Gazdzicki:1998vd}
before its observation was based on the assumption of a first-order phase transition with the formation of a mixed phase. 
In a refined formulation within a Boltzmann transport approach the horn behaviour was found and could be associated with a first-order quark-hadron phase transition \cite{Nayak:2010uq}.
In view of the ab-initio LQCD result that the transition under these conditions rather is a crossover \cite{Bazavov:2018mes,Borsanyi:2020fev},
a sound theoretical explanation of the effect is still under scrutiny.
Moreover, the recent data from the NA61 experiment have shown a strong dependence on the system size with no horn effect in Ar+Sc collisions
\cite{Gazdzicki:2020jte} for which the reason is not yet understood.
In another, purely hadronic transport model, the basic pattern of the energy scan of strange-to-nonstrange particle production ratios including the $K^+/\pi^+$  and the $\Lambda/\pi^-$ horns could be reproduced because the lifetime of the (hadronic) fireball until freeze-out gets shortened when increasing the energy of the collision above $30$ AGeV \cite{Tomasik:2006qs}.  
A rather successful theoretical description of the  experimental data could be given within yet another transport approach, the parton-hadron-string-dynamics (PHSD) approach \cite{Palmese:2016rtq,Moreau:2017dgq} where strangeness production in the hadronic phase was enhanced by a modification of the 
Schwinger production rates due to partial chiral symmetry restoration. 
This regime was abruptly terminated when the system evolved through a partonic phase at energies above $\sqrt{s_{NN}}\sim 7-10$ GeV, resulting in a rather sharp drop of the $K^+/\pi^+$ ratio as a result of the chiral condensate melting and QGP formation during the collision.
This microscopic transport model also predicts a smoothing of the peak with decreasing system size \cite{Moreau:2017dgq} albeit not quite matching the behaviour recently observed by NA61 in Ar+Sc collisions \cite{Gazdzicki:2020jte}. 

In the present work we want to elaborate on the description of the kaon-to-pion ratios within a nonequilibrium statistical model where the interplay of strangeness enhancement and quick approach to the QGP asymptotics is regulated by quasi-chemical potentials along the chemical freeze-out line in the QCD phase diagram that is parametrized as in Ref. \cite{Cleymans:2005xv}. 
In this description the pion quasi-chemical potential (which is the generalization of the chemical potential to the nonequilibrium case) plays a special role. 
According to the Zubarev approach to nonequilibrium statistical systems such a quasi-chemical potential appears as a Lagrange multiplier in a generalized Gibbs ensemble describing the nonequilibrium situation of a quasi-conserved pion number. 
While pions can be quickly created in the early phase of the heavy-ion collision, for instance by the Schwinger mechanism in a strong color flux tube field 
\cite{Dyrek:1990ec,Florkowski:2003mm} or by  the conversion of a gluon saturated initial state (color glass condensate) to an oversaturated pion gas, e.g., by processes of the kind $g+g\to \pi+\pi$ \cite{Nazarova:2019dif}, their number is quasi conserved over the time scales of a heavy-ion collision until freeze-out,
since their electroweak ($\pi^\pm \to \mu^\pm + \nu_\mu$) or two-photon ($\pi^0\to 2 \gamma$) decays are several orders of magnitude slower.  
Strong pion absorption processes ($\pi+\pi \to \rho$, $\pi+N\to \Delta$) are fast but do not change the net pion number since these resonances decay on the 
way to the detector or even during the fireball lifetime.
Therefore, the adequate approach to this nonequilibrium situation is the Zubarev approach of the generalized Gibbs ensemble where the quasi conserved pion number requires an additional Lagrange multiplier, the pion chemical potential $\mu_\pi$, as an additional parameter of the statistical operator.
For details, see \cite{Blaschke:2020afk}.  
The resulting nonequilibrium pion distribution function with the pion chemical potential has been successfully used in the phenomenological description of the transverse momentum spectra of pions produced in heavy-ion collision experiments, see \cite{Kataja:1990tp,Begun:2013nga}.

This paper is organized as follows.
In Sec.~\ref{sec:1}, the formalism of the PNJL model with vector interaction and entanglement (EPNJL model) is introduced and the structure of the phase diagram 
is discussed. 
The behavior of mesons and quarks at zero and finite density is discussed in Sec.~\ref{sec:2} both for the pole approximation to the Bethe-Salpeter equation and for the Beth-Uhlenbeck approach. 
In Sec.~\ref{sec:Kpi} the results for the $K^+/\pi^+$ and $K^-/\pi^-$  ratios and the lessons for the phenomenology of matter created in heavy-ion collisions will 
be discussed.
In Sec.~\ref{sec:concl} we present our Conclusions.

\section{The SU(3) PNJL and EPNJL models}
\label{sec:1}

In this work we start from the  SU(3) PNJL model with the Kobayashi - Maskawa - t'Hooft (KMT) interaction \cite{Fukushima:2008wg,Bratovic:2012qs,Chu:2016ixv}
\begin{eqnarray}
\mathcal{L\,} & = & \bar{q}\,(\,i\,{\gamma}^{\mu}\,D_{\mu}\,-\,\hat
{m} - \gamma_0\hat{\mu})\,q +  \frac{1}{2}\,g_{S}\,\,\sum_{a=0}^{8}\,[\,{(\,\bar{q}\,\lambda
^{a}\,q\,)}^{2}\,\,+\,\,{(\,\bar{q}\,i\,\gamma_{5}\,\lambda^{a}\,q\,)}%
^{2}\,] + \nonumber \\
&+&  g_D \,\,\{\mbox{det}\,[\bar{q}\,(\,1\,+\,\gamma_{5}%
\,)\,q\,]+\mbox{det}\,[\bar{q}\,(\,1\,-\,\gamma_{5}\,)\,q\,]\,\} -  \mathcal{U}(\Phi, \bar{\Phi}; T), 
\label{lagr}%
\end{eqnarray}
where $q$ is the quark field with three flavours $(u,d,s)$, $\lambda^{a}$  are the Gell-Mann matrices used here in flavor space, the diagonal matrices   
$\hat{m}=\mbox{diag}(m_{u},m_{d},m_{s})$ and $\hat{\mu}=\mbox{diag}(\mu_{u},\mu_{d},\mu_{s})$ contain the current quark masses and quark chemical potentials, respectively. 
The covariant derivative D$_\mu = \partial^\mu -i A^\mu$ contains the gauge field $A^\mu(x) = g_S A^\mu_a\frac{\lambda_a}{2}$ in a notation which absorbs the strong interaction coupling.
In the Polyakov-loop extended NJL model one uses $A^\mu(x)=\delta_{\mu,0}A^0$ with $A^0= -  iA_4$ being a homogeneous background field with only the color diagonal fields  $A^0_3$ and $A^0_8$ having nonvanishing values. 
The confinement properties are described by the effective potential $\mathcal{U}(\Phi, \bar{\Phi}; T)$, which depends on the complex traced Polyakov loop
$\Phi={\rm Tr}_c \exp[i\beta (\lambda_3 A_3^0+\lambda_8A_8^0)]$ and its conjugate $\bar{\Phi}$.
The potential is constructed on the basis of the center symmetry $Z_3$ of the color SU(3) gauge group and its Haar measure, see also \cite{Sasaki:2012bi}. 
The possible temperature dependence of its parameters is fitted to lattice QCD results for the pressure in the pure gauge sector (for details see \cite{Ratti:2005jh,Blanquier:2011zz,Friesen:2011wt,Dubinin:2016wvt}). 
In this work, we use the standard polynomial form of the effective potential $\mathcal{U}(\Phi, \bar{\Phi}; T)$ \cite{Ratti:2005jh,Friesen:2011wt} where the 
original parameter $T_0=270$ MeV for the deconfinement transition temperature in pure gauge is changed to $T_0=187$ MeV for the system with 
$N_f=2+1$ flavors \cite{Schaefer:2007pw}.

The grand thermodynamic potential density $\Omega(T,\{\mu_i\})$ can be obtained directly from the Lagrangian density (\ref{lagr}) when the mean-field approximation is applied. Then, the following gap equations for the constituent quark masses are obtained by the stationarity conditions  
$\partial \Omega/\partial m_i=0$ as
\begin{eqnarray}
m_i = m_{0i} - 2 g_S\langle\bar{q_i}q_i\rangle - 2g_D\langle\bar{q_j}q_j\rangle\langle\bar{q_k}q_k\rangle, 
\label{eqgap}  
\end{eqnarray}
where $i, j, k = $u, d, s are in cyclic order and the quark condensates are defined as
\begin{eqnarray}
\langle\bar{q_i}q_i\rangle &=& \frac{\partial  \Omega}{\partial m_{0i}} = - 2 N_c \int\frac{d^3p}{(2\pi)^3}\frac{m_i}{E_i}(1 - f^+_\Phi(E_i) - f^-_\Phi(E_i)),
\end{eqnarray}
with $E_{i}=\sqrt{{p}^2+m_{i}^2}$ being the energy dispersion relation for a quark of flavor $i$.
From the thermodynamic potential density other thermodynamic quantities of interest can be obtained by derivation like, e.g., the quark number densities
\begin{eqnarray}
n_i & = & \frac{\partial \Omega}{\partial \mu_i} = 2N_c\int\frac{d^3p}{(2\pi)^3}(f^+_\Phi(E_i) - f^-_\Phi(E_i))~.
\end{eqnarray}
The quark condensates and the quark densities at finite temperature are defined with  the modified Fermi functions $f^\pm_\Phi(E_i)$ of the PNJL model
\begin{eqnarray}
\label{f-Phi}
f^+_\Phi(E_f)=
\frac{(\bar{\Phi}+2{\Phi}Y)Y+Y^3}{1+3(\bar{\Phi}+{\Phi}Y)Y+Y^3}
~, \ \ \ f^-_\Phi(E_f)=
\frac{({\Phi}+2\bar{\Phi}\bar{Y})\bar{Y}+\bar{Y}^3}{1+3({\Phi}+\bar{\Phi}\bar{Y})\bar{Y}+\bar{Y}^3}~,
\label{f-Phi-bar}
\end{eqnarray}
which are obtained after Matsubara summation when the gluon background field is accounted for \cite{Ratti:2005jh}. 
Here the abbreviations $Y={\rm e}^{-(E_f-\mu_f)/T}$ and $\bar{Y}={\rm e}^{-(E_f+{\mu_f})/T}$ are used. 
The traced Polyakov loop variables as order parameters of deconfinement have the two limits $\Phi=\bar \Phi=0$ in the confined phase and $\Phi=\bar \Phi=1$
in the deconfined phase.
In these limits, the functions (\ref{f-Phi-bar}) go over to ordinary Fermi functions,
\begin{eqnarray}
\label{f-Phi-0}
f^\pm_{\Phi=0}(E_f)=[{\rm e}^{3(E_f\mp \mu_f)/T}+1]^{-1}~, \ \ \ f^\pm_{\Phi=1}(E_f)=[{\rm e}^{(E_f\mp \mu_f)/T}+1]^{-1}~,
\label{f-Phi-1}
\end{eqnarray}
where the effect of confinement on the quark distribution function consists in a rescaling of the temperature $T\to T/3$, when compared to the deconfined case.

\begin{figure}
\centerline{
\resizebox{0.45\columnwidth}{!}{%
\includegraphics{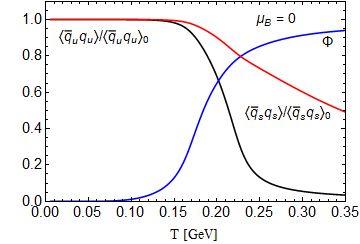}}
\resizebox{0.45\columnwidth}{!}{%
\includegraphics{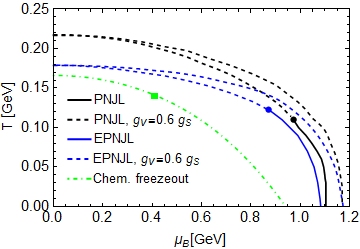}}}
\caption{The normalized chiral condensate for u- and s- quarks and the traced Polyakov loop $\Phi$ as functions of the temperature (left panel). 
Phase diagram in the T-$\mu_B$ plane (right panel) indicating the pseudocritical temperatures of the chiral crossover transition for the PNJL (black dashed lines) and the EPNJL (blue dashed lines) models in the case of nonzero vector coupling ($g_V=0.6\, g_S$). For vanishing vector coupling the crossover merges a first 
order transition for both cases (solid lines) in a critical endpoint (filled dots). For orientation, the chemical freezeout line parametrized in \cite{Cleymans:2005xv} is
shown as a green dash-dotted line with a filled square on it indicating the freezeout of heavy-ion collisions with $\sqrt{s_{NN}}=8$ GeV, where the peak in the $K^+/\pi^+$ ratio is situated.}
\label{3pd}
\end{figure}

Due to their coupling to the chiral condensate, the quarks develop a quasiparticle mass even for vanishing current quark masses $m_{0i}$ (chiral limit).
This phenomenon is called dynamical chiral symmetry breaking. 
Simultaneously, the quarks are coupled to the homogeneous gluon background fields representing the Polyakov loop dynamics. 
At low temperature, the chiral symmetry is spontaneously broken due to the finite value of the quark condensate and confinement is observed in the system 
($\Phi\rightarrow 0$).  
Chiral symmetry is restored when the dynamically generated quark mass drops as a function of temperature and chemical potentials (see Fig.~\ref{3pd}, left panel). 
For finite temperatures, the Polyakov field $\Phi$ becomes nonzero in a crossover type transition, which entails $Z_3$-symmetry breaking. 
The point where the crossover changes to a first order transition is the critical end point (CEP) (see Fig.\ref{3pd}, right panel). 
At low temperature and high chemical potential, below the CEP, the thermodynamic potential density has three extrema corresponding to three solutions of 
the gap equations (\ref{eqgap}). Two of them are minima with a maximum in-between them. 
The situation when the minima have equal depth corresponds to equal pressures in the corresponding phases. When changing the thermodynamic parameters 
($T$ and/or $\mu$) the system goes over from one to the other minimum via the barrier, resulting in a jump of the order parameter (mass or chiral condensate) 
that signals a first order phase transition. 

For the numerical calculations in this work we used the following set of parameters: 
the current quark masses $m_{0 u} = m_{0 d} = 5.5$ MeV, $m_{0 s} = 0.131$ GeV,  the cut-off $\Lambda = 0.652$ GeV and
the couplings 
$g_D \Lambda^5 = 10.592$, $g_S \Lambda^2 = 1.828$. 
We also introduced the strange quark chemical potential as $\mu_s = 0.55\mu_u $ and $\mu_u = \mu_d$ \cite{Friesen:2018ojv}. 
It can be seen from the right panel of Fig.~\ref{3pd} that the pseudocritical temperature of the chiral crossover transition at $\mu_B = 0$ GeV 
in the PNJL model case is $T_c = 0.218$ GeV (black dashed line), and thus higher than the $T_c =156.5\pm 1.5$ GeV  obtained  by the ab-initio simulations of
Lattice QCD \cite{Bazavov:2018mes}. 

It could not be expected that the PNJL quark matter model in the mean field approximation would yield the chiral restoration temperature in accordance with the Lattice QCD result. 
Namely, while in the NJL model due to lack of confinement the excitation of quarks at low temperatures mimics the presence of a thermal medium,
because of quark confinement (more precisely: statistical suppression by the Polyakov-loop) in the PNJL model at mean field level even those excitations are missing. 
In nature, however, the thermodynamics at low temperatures is dominated by hadronic excitations (pions, kaons, ...) which in turn play the dominant role in melting the chiral condensate.  
This has been demonstrated by calculating the chiral condensate within the hadron resonance gas model, e.g., in Ref.~\cite{Jankowski:2012ms}. 
The formation of hadrons as bound states of quarks and in particular their backreaction on the chiral condensate is, however, not accounted for 
in the mean field approximation.  

An improvement of the correspondence between a selfconsistent mean field calculation of the chiral condensate and Polyakov-loop on the one hand and their behaviour in Lattice QCD has been obtained within the so-called Entanglement PNJL model \cite{Ruivo:2012xt,Sugano:2014pxa}.
According to this extension of the PNJL model, a phenomenological dependence of the scalar meson coupling $g_S$ on the Polyakov loop is introduced that does obey the $Z_3$-symmetry 
\begin{equation}
\tilde{g}_S(\Phi) = g_S(1-\alpha_1\Phi\bar{\Phi} - \alpha_2(\Phi^3 + \bar{\Phi}^3)),
\label{Gt}
\end{equation}
where the parameters $\alpha_1 = \alpha_2 = 0.2$ were chosen in \cite{Sakai:2010rp} to reproduce the two-flavor LQCD data. 
As soon as the  traced Polyakov-loop deviates from zero $g_S$ gets   reduced which in turn lowers the quark mass and the chiral condensate.
Such rescaling of $g_S$ leads to a rescaling of the pseudocritical temperature in the low-density region \cite{Sakai:2010rp,Ruivo:2012xt,Kalinovsky:2016dik}
to $T_c\sim 0.179$ GeV, see the blue lines in the right panel of Fig.~\ref{3pd}, now being synchronous with the Polyakov-loop transition. 

A better solution should be possible within a PNJL-like model when going beyond the mean field level of desciption and considering the influence of the
hadronic excitations in the medium on the chiral condensate and the quark selfenergies (see, e.g.,  \cite{Radzhabov:2010dd,Blaschke:2017pvh,Blaschke:1995gr,Kitazawa:2014sga}), 
but such an approach would still have to be developed to include a full hadron resonance gas as a limiting case.

The position of  the CEP 
is under debate \cite{Stephanov:2007fk} and can not be settled yet by Lattice QCD simulations. 
Actually, there are several possibilities for the topology of the QCD phase diagram besides the simple alternative with one CEP or crossover all over; 
for a short discussion see \cite{Contrera:2016rqj}.
It was shown in the NJL \cite{Sasaki:2006ws} and in the PNJL model \cite{Bratovic:2012qs} with vector interaction, that the CEP can disappear when the vector coupling exceeds a critical value. 
The Lagrangian of the PNJL model with vector interaction is obtained from the Lagrangian (\ref{lagr}) by adding the vector interaction contribution
\begin{eqnarray}
\mathcal{L}_{\rm V}  =  - \frac{1}{2}g_{V} \sum_{a=0}^{8}\,\,{(\,\bar{q}\gamma_\mu\lambda^{a}\,q\,)}^{2}.
\label{lagr_v}%
\end{eqnarray}
In this case the set of the equations of motion will be extended by the equations for the vector mean fields which are absorbed in the rescaled chemical potentials 
$\tilde{\mu_i}$ that replace the chemical potentials in the distribution functions (\ref{f-Phi-bar}), where 
\begin{eqnarray}
\tilde{\mu}_i = \mu_{i} - 2 g_{V} n_i.
\end{eqnarray}
As a matter of fact, the vector coupling can also be rescaled in the spirit of the EPNJL model, similar to the scalar coupling  (\ref{Gt}),
\begin{equation}
\tilde{g}_V(\Phi) = g_V(1-\alpha_1\Phi\bar{\Phi} - \alpha_2(\Phi^3 + \bar{\Phi}^3)),
\label{Gvt}
\end{equation}
with the same parameters as in Eq.~(\ref{Gt}). For results see, e.g., \cite{Friesen:2014mha,Sugano:2014pxa}.
In the present work we will consider the case when only $g_S$ depends on the medium via the traced Polyakov-loop variables 
$\Phi$ and $\bar{\Phi}$. 
In Fig. \ref{3pd} we show that for $g_V=0.6~g_S$ the first order transition domain and the CEP disappear 
from the phase diagram and only a crossover takes place (black and blue dashed lines for PNJL and EPNJL models, respectively).

Despite the principal problem in explaining the phase structure of QCD within chiral quark models like PNJL and EPNJL at the 
mean field level (i.e. by ignoring the effects of hadronic excitations on the order parameters) we find by comparing with the lattice QCD 
data \cite{Bazavov:2018mes} that the result for the EPNJL model provides an acceptable proxy for the phase border where the 
hadronization transition shall take place in the dynamical evolution of the QGP.
It is reassuring for the application of the EPNJL model to heavy-ion collisions at the end of this contribution that the parametrized 
line of the chemical freezeout \cite{Cleymans:2005xv} shown as a green line in the right panel of Fig. \ref{3pd} lies entirely in the 
confining phase with broken chiral symmetry, but also sufficiently close to the hadronization transition.
The basis for such applications, in particular to the celebrated kaon-to-pion ratios and the horn effect for the energy dependence of 
$K^+/\pi^+$, is a field theoretical description of scalar and pseudoscalar meson bound states and their dissociation 
by the Mott effect which occurs under extreme conditions of temperature and density. 
This we will describe in the next section.

\section{Mesons in hot and dense matter}
\label{sec:2}

In this section we will outline the field-theoretical approach to the description of scalar and pseudoscalar bound states and their 
Mott dissociation  in a hot and dense medium within a chiral quark model of the PNJL type. 
The underlying calculations are developed more in detail in \cite{Dubinin:2016wvt,Kalinovsky:2016dik} and references therein.

The meson properties such as masses and decay widths are encoded in the quark-antiquark scattering amplitude (T-matrix) that shares 
the analytic properties with meson propagator ($M=P,S$)
\begin{equation}
\label{eq:SM}
\mathcal{S}_{ij}^M (\omega,\mathbf{q}) = \frac{2 P_{ij}}{1 - 4 P_{ij}\Pi_{ij}^{M}(\omega,\mathbf{q})}~, \label{rprop}
\end{equation}
where in the random phase approximation (RPA) the polarization function for pseudo-scalar mesons ($M=P$) at rest ($\mathbf{q=0}$) 
in the medium has the form
\begin{equation}
\Pi_{ij}^{P}(\omega, \mathbf{0})=4\left\{I_{1}^{i}(T,\mu_i)+I_{1}^{j}(T,\mu_j) - [(\omega+\mu_{ij})^{2} - (m_{i}-m_{j})^{2}]\, I_{2}^{ij}(\omega; T, \mu_{ij})\right\},
\nonumber\\
\label{ppij}
\end{equation}
where $\mu_{ij}=\mu_i-\mu_j$ is the meson chemical potential and for $P=\pi,K$ we have \cite{Rehberg:1995kh}
\begin{eqnarray}
P_{ij}^{\pi}&=&g_{S}+g_{D}\left\langle\bar{q}_{s}q_{s}\right\rangle, \label{Ppi} \\
P_{ij}^{K} &=& g_{S}+g_{D}\left\langle\bar{q}_{u}q_{u}\right\rangle. \label{Pkaon}
\end{eqnarray}
The first integral  $I_{1}^{i}(T,\mu_i)$ is a one-loop integral that does not depend on the energy $\omega$ and after Matsubara summation takes the form
\begin{eqnarray}
\label{i1}
I_1^i(T,\mu_i) &=& \frac{N_c}{4\pi^2} \int_0^\Lambda \frac{dp \, p^2}{E_i} \left[1-f^-_\Phi(E_i) - f^+_\Phi(E_i) \right].
\end{eqnarray}
The second one depends (after analytic continuation) on the complex energy variable $z$ and can be decomposed into two contributions \cite{Yamazaki:2013yua}
\begin{eqnarray}
\label{i2}
I_2^{ij} (z,T,\mu_{ij})&=&I_{\rm 2,pair}^{ij} (z,T,\mu_{ij})+I_{\rm 2,scatt}^{ij} (z,T,\mu_{ij})~,
\end{eqnarray}
where the first one, 
\begin{eqnarray}
\label{I2-1}
I_{\rm 2,pair}^{ij} (z,T,\mu_{ij}) &=& \frac{N_c}{8\pi^2} \int_0^\Lambda \frac{dp \, p^2}{E_iE_{j}}
\Biggl[ \frac{1-f^+_\Phi(E_{j})-f^-_\Phi(E_i)}{z-E_i-E_{j}-\mu_{ij}} 
-  \frac{1-f^+_\Phi(E_{i})-f^-_\Phi(E_{j})}{z+E_i+E_{j}-\mu_{ij}} 
\Biggr],\nonumber \\
\end{eqnarray}
exhibits poles in the integrand which correspond to pair excitation modes (i.e., with the sum of the one-particle energies in the denominators) 
and the second one,
\begin{eqnarray}
\label{I2-2}
I_{\rm 2,scatt}^{ij} (z,T,\mu_{ij})&=& \frac{N_c}{8\pi^2} \int_0^\Lambda \frac{dp \, p^2}{E_iE_{j}}
\Biggl[\frac{f^+_\Phi(E_{j})-f^+_\Phi(E_{i})}{z+E_{i}-E_{j}-\mu_{ij}} 
- \frac{f^-_\Phi(E_{j})-f^-_\Phi(E_{i})}{z-E_{i}+E_{j}-\mu_{ij}} 
\Biggr], 
\end{eqnarray}
has poles in the integrand which correspond to scattering modes (i.e., with the difference of the one-particle energies in the denominators), see also \cite{Blaschke:2013zaa,Dubinin:2016wvt,Yamazaki:2013yua}.
Generally, the meson propagator (\ref{eq:SM}) can have poles and cuts in the $z$- plane, whereby the poles on the real axis correspond to stable bound states, 
and a cut stands for the continuous spectrum of scattering states. 
Unstable states with a short lifetime, like resonances in the scattering state continuum can be treated as complex poles with an imaginary part corresponding to inverse lifetime of the state. In the following we will sketch two ways of dealing with the analytic properties of the meson propagator and to arrive at results for the thermodynamics of quark matter with mesonic correlations: i) the Breit-Wigner approximation that approximates mesonic correlations as complex poles near the real $z$- axis and ii) the Beth-Uhlenbeck approach that represents the analytic properties along the real axis by a phase shift function that plays the role of a spectral density for the pressure and thus the thermodynamic potential.
 
 \subsection{Breit-Wigner approximation}
 
\begin{figure}[h]
\centerline{
\resizebox{0.48\columnwidth}{!}{%
\includegraphics{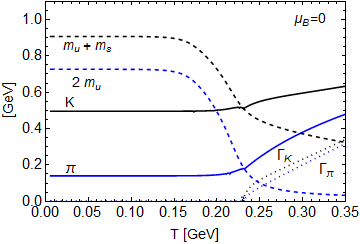}}
\resizebox{0.46\columnwidth}{!}{%
\includegraphics{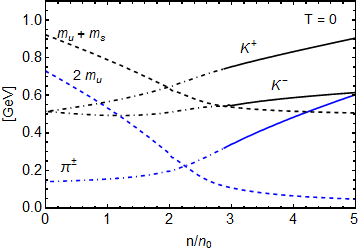}}}
\caption{Meson masses (solid lines) for the PNJL model as functions of the temperature at $\mu_B = 0$ (left panel) and as functions of the 
normalized baryon number density at $T=0$ (right panel), together with the threshold masses of the corresponding quark-antiquark continuum (dashed lines). 
Note that between $n=0$ and $n=2.9\, n_0$, there is no stable solution for the baryon density and 
the meson masses are indicated by dash-dotted lines to guide the eye.}
\label{masses}
\end{figure}

 In the Breit-Wigner approximation, one is looking for a complex pole $z_M=m_M - i \Gamma_M/2$ of the meson propagator, which is a zero of its denominator 
 function 
 \begin{equation}
1 - 2 P_{ij}\Pi_{ij}^{M}(z=z_M,\mathbf{q=0}) =0, 
\label{rdisp}
\end{equation}
for which $\Gamma_M \ll m_M$ should hold.
Since in general the polarization function is complex, Eq. (\ref{rdisp}) defines a system of two real equations for the mass $m_M$ and the decay width 
$\Gamma_M$ of the mesonic state $M$, see Appendix A of Ref.~\cite{Dubinin:2016wvt}. 
When the mass parameter is below the two-quark threshold ($m_M < m_i + m_j$), the polarization function is real and Eq. (\ref{rdisp}) corresponds to the 
homogeneous Bethe-Salpeter equation \cite{Blanquier:2011zz,Rehberg:1995kh} in the rest frame of the meson which defines a true bound state with 
mass $m_M$ and infinite lifetime ($\Gamma_M=0$).
In the left panel of Fig. \ref{masses}, we show the temperature dependences of the masses of the pseudoscalar mesons $\pi$ and $K$ and the threshold 
masses of their corresponding two-quark continuum. 
As can be seen, when the mass of the meson exceeds the sum of the masses of its constituents ($m_M > m_i + m_j$), the meson turns into a resonance 
state in the continuum with a finite lifetime ($\Gamma_M > 0$) and the Mott transition occurs. 
In this case, the complex properties of the integral $I_2$ in Eq. (\ref{i2}) have to be taken into account and the mass pole solution of the Bethe-Salpeter equation 
moves away from the real axis into the complex $z$- plane. 
The temperatures of the Mott transition for pion and kaon are $T^\pi_{\rm Mott} = 0.232$ GeV and $T^K_{\rm Mott} = 0.230$ GeV, respectively. 
The solution of the Bethe-Salpeter equation (\ref{rdisp}) can also be performed at finite baryon chemical potential $\mu_B\neq 0$.
Results for the meson masses and corresponding two-quark continuum thresholds are shown in the right panel of Fig.~\ref{masses} as a function of the baryon number density in units of the nuclear saturation density $n_0=0.15$ fm$^{-3}$. 
Note that between $n=0$ and $n_c=2.9\,n_0$, there is no stable solution for the baryon density, since the Maxwell construction of the first-order phase transition 
in the absence of a nuclear matter phase joins the quark matter phase at $n=n_c$ directly with vacuum state at $n=0$.

\subsection{Beth-Uhlenbeck approach}

\begin{figure}[t]
\centerline{
\resizebox{0.48\columnwidth}{!}{
		\includegraphics{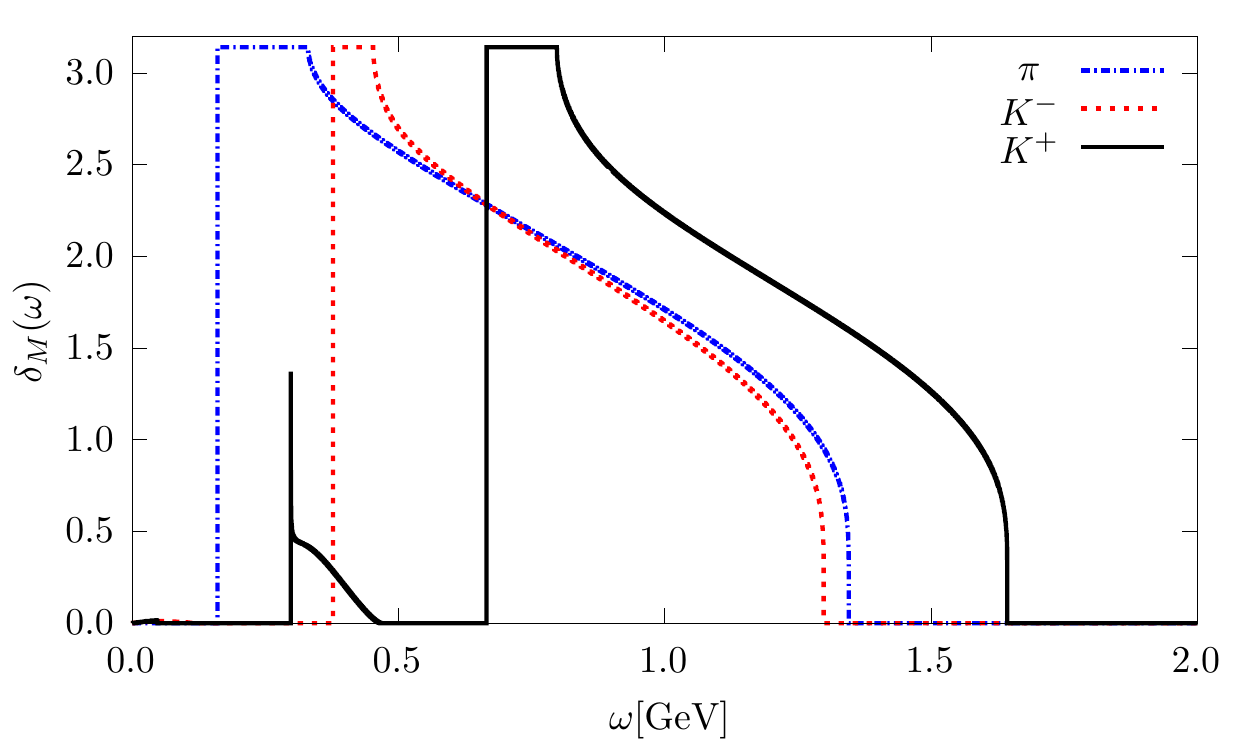}}
\resizebox{0.48\columnwidth}{!}{
		\includegraphics{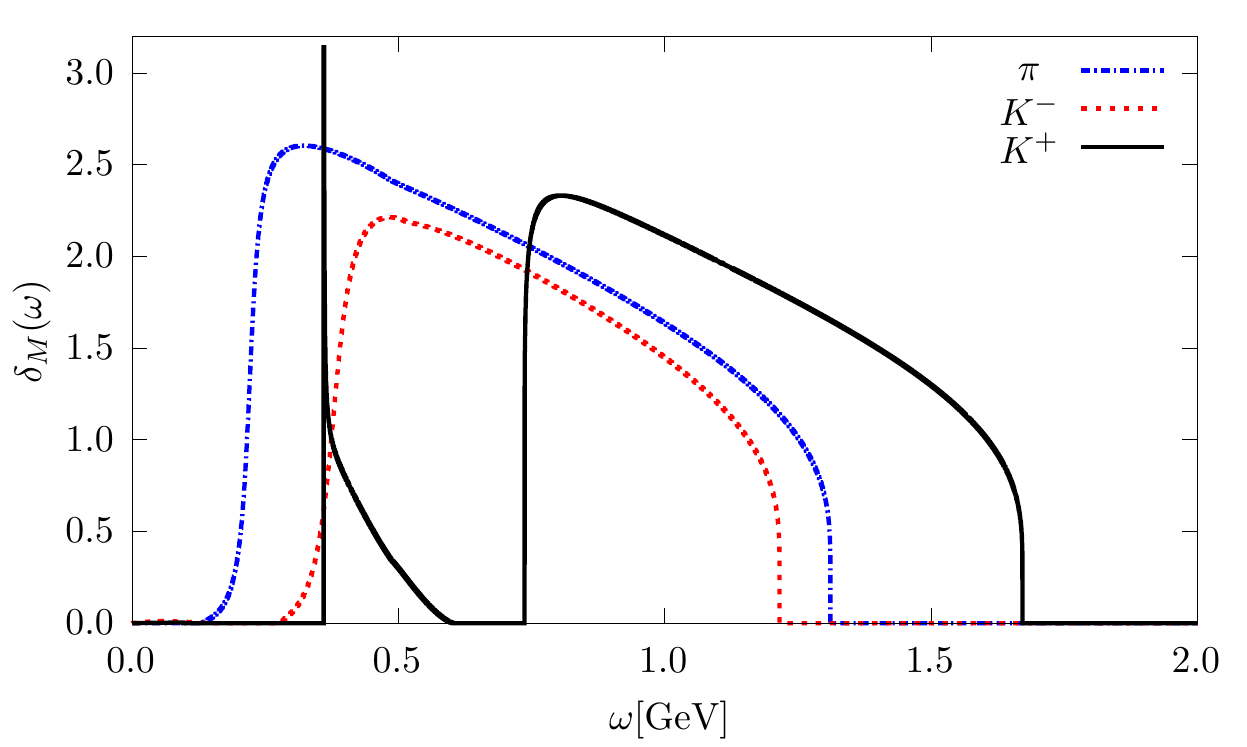}}}
	\caption{Dependence of the mesonic phase shift in the PNJL model case  for the pion (blue dash-dotted line), the $K^-$ (red dotted line) and the $K^+$ (black solid line) channels on the center of mass energy $\omega$ for $T= 193$ MeV and nonstrange chemical potential $\mu_u=181$ MeV  (i.e. for $T/\mu_B=0.36$, left panel) and $\mu_u=240$ MeV at the same temperature (i.e. for $T/\mu_B=0.27$, right panel). 
The phase shift for $K^+$ contains an anomalous mode at energies below the "normal" mode (left panel) which eventually becomes a true bound state (plasmon pole) in the deconfined phase (right panel), when the normal mode of $K^+$ underwent a Mott dissociation and turned into a resonant correlation in the continuum, like also $\pi$ and $K^-$. }
	\label{Fig:ShiftsPiKaMiKaPl}%
\end{figure}

The RPA for the meson polarization function (\ref{ppij}) provides a much richer analytic structure than just a complex mass pole near the real $z$- axis.
Under given conditions for temperature and chemical potential, from the four denominators in the pairing and scattering  channels of the $I_2$-integral 
(\ref{i2}) can simultaneously emerge poles for the bound state and cuts for the continuum states which cannot adequately be represented by a single complex 
mass pole approximation. 
In such a case it is advantageous to introduce the phase shift in the corresponding channel and to use the Beth-Uhlenbeck formula for evaluating the 
contribution of the correlations in that channel to the thermodynamic potential.  
To this end, the complex mesonic propagator in Eq.~(\ref{rprop}) can be rewritten in the polar representation with a modulus and a phase
\begin{equation}
\mathcal{S}^{M}_{ij}(\omega,\mathbf{q})= |\mathcal{S}^{M}_{ij}(\omega,\mathbf{q})|\, {\rm e}^{i \delta_M(\omega,\mathbf{q})},
\end{equation}
where the meson phase shift is defined as
\begin{equation}
\delta_M(\omega,\mathbf{q}) = -{\rm{arctan}}\left\{\frac{{\rm{Im}}[{\mathcal{S}}^{M}_{ij}(\omega+i\eta,\mathbf{q})]}{{\rm{Re}}[\mathcal{S}^{M}_{ij}(\omega+i\eta,\mathbf{q})]}\right\}.
\label{phaseshift}
\end{equation}
According to Levinson's theorem, the bound state appears at the energy, where the phase shift jumps up by the value $\pi$ \cite{Wergieluk:2012gd}. 
In the rest frame of the meson this energy corresponds to the meson mass. 
In Fig.~\ref{Fig:ShiftsPiKaMiKaPl} the results for the phase shifts of the pion and the charged kaons are shown for the PNJL model case at a temperature 
$T= 193$ MeV and a nonstrange chemical potential $\mu_u=181$ MeV (left panel) and $\mu_u=240$ MeV (right panel). 
According to the phase diagram of Fig.~\ref{3pd} (right panel) this point is situated on the hadronic side of the phase border in the $T-\mu_B$ plane where the 
where the mesons are true bound states, i.e. where $T$ is below the corresponding Mott temperatures.
Correspondingly, one can read off the masses of the meson bound states from the positions where the phase shift jumps by $\pi$.
One recognizes the mass splitting of the charged kaon states so that $K^-$ is lighter than $K^+$ at finite baryon density, as in the right panel of 
Fig.~\ref{masses}. 
But from the phase shift one  can get more information, namely about the nonvanishing, negative contribution of the continuum of scattering states to the thermodynamics ($d\delta_M/d\omega<0$) and about the "anomalous" scattering mode  that develops below the bound state in the 
"normal" pairing mode of the propagator  in the $K^+$ channel.
When crossing the phase border to the deconfined phase, the normal mode undergoes a Mott dissociation and turns into a resonant correlation in the continuum while the anomalous mode can even become a true bound state (plasmon pole), see the right panel of Fig.~\ref{Fig:ShiftsPiKaMiKaPl} and 
Ref.~\cite{Dubinin:2016wvt}.

Both Fig.~\ref{masses} and Fig.~\ref{Fig:ShiftsPiKaMiKaPl} clearly demonstrate the splitting of the charged multiplet mass with increasing density. 
The splitting can be explained by the modification of the Fermi sea by the presence of the medium \cite{Friesen:2018ojv,Lutz:1994cf,Costa:2003uu}. 
In dense baryon matter the concentration of light quarks is very high  
so that the Pauli blocking factor $Q_{ij}=1-f^+_\Phi(E_{j})-f^-_\Phi(E_i)$ in the $I_{\rm 2,pair}^{ij} $ integral (\ref{I2-1}) for $i=u,d$ being a light quark flavor and
$j=s$ (the case of $K^+$ and $K^0$) reduces the integrand at low one-particle energies, viz. momenta, and thus shifts the bound state pole to higher values of the energy $z$ where the Pauli blocking is less prominent. 
Finally, when the bound state pole reaches the continuum threshold, $z_M-\mu_{ij}=E_i(p=0)+E_j(p=0)= m_i+m_j$, the binding energy vanishes and this 
marks the condition for the Mott dissociation of the bound state. 
At the same time, with increasing density of light quark states, the difference between light and strange quark occupation numbers increases so that the 
integral $I_{\rm 2,scatt}^{ij}$ for the scattering mode can become dominant in the polarization function (\ref{ppij}) and induce a bound state pole in the meson propagator (\ref{eq:SM}), see the right panel of Fig.~\ref{Fig:ShiftsPiKaMiKaPl}.  

In the case of  $K^-$ and $\bar{K}^0$, the meson is composed of a strange quark and light antiquark which both are not as abundant in the baryon-rich medium as the light quarks, so that the Pauli blocking does not operate (the normal meson mode has an approximately medium-independent mass) and the anomalous mode cannot develop since the phase space occupation factors cancel each other.
For more details see \cite{Dubinin:2016wvt}.

\section{"Horn" effect in the kaon-to-pion ratio}
\label{sec:Kpi}

In this Section we will discuss the application of the PNJL-type models to a description of the kaon-to-pion ratios that are measured in heavy-ion collision experiments. 
The peak (also called the "horn"), that has been found in the ratio of positively charged kaons to pions ($K^+/\pi^+$) at energies $\sqrt{s_{NN}}\sim$ 7-10 GeV 
by the NA49 experiment at CERN SPS in an energy scan with fixed-target Pb+Pb collisions and that becomes apparent in the world data on this ratio for large colliding systems (Pb+Pb and Au+Au, see Fig. 1 of \cite{Aduszkiewicz:2019zsv})  has been 
discussed as a signal of QGP formation \cite{Afanasiev:2002mx,Alt:2007aa,Adamczyk:2017iwn}. 
The sudden drop in the ratio after reaching maximum at $\sqrt{s_{NN}}\sim$ 8 GeV could be explained as a result of QGP formation during the collision \cite{Andronic:2008gu,Gazdzicki:1998vd,Nayak:2010uq,Palmese:2016rtq} although there are purely hadronic kinetic models that can describe the "horn" 
feature too \cite{Tomasik:2006qs}.
Generally, the yields and ratios of hadrons (including light nuclear clusters) produced in heavy-ion collisions are described astonishingly well within the thermal statistical model of hadron resonances, where conservation laws of baryon number, isospin and strangeness are implemented, and the sudden (chemical) freezeout is followed by resonance decays.  
Within this approach, the abundances of all identified hadrons are simultaneously described by just two thermodynamic variables: the temperature $T$ and baryochemical potential $\mu_B$ at freezeout.
They vary as a function of the collision energy  $\sqrt{s_{NN}}$	in a systematic way \cite{Andronic:2017pug} that can be parametrized in a simple functional form,
with typical forms given in \cite{Andronic:2005yp} and in \cite{Cleymans:2005xv}, the latter being
\begin{eqnarray}
T(\mu_B) &=& a - b\mu^2_B- c\mu^4_B~,~~\mu_B(\sqrt{s}) = \frac{d}{1 + e\sqrt{s}},
\label{param_exp1}
\end{eqnarray}
where $a = 0.166 \pm 0.002$ GeV, $b = 0.139 \pm 0.016$ GeV$^{-1}$,
and $c = 0.053 \pm 0.021$ GeV$^{-3}$, $d = 1.308 \pm 0.028$ GeV, $e = 0.273 \pm 0.008$ GeV$^{-1}$.  
The freezeout parameters form a line in the QCD phase diagram (see the green dash-dotted line in the right panel of Fig.~\ref{3pd}) along which one can  
consider the $K^+/\pi^+$ ratio according to the thermal statistical model and compare this curve with the experimental data.
This has been done in \cite{Andronic:2005yp} and a very broad peak at the position of the "horn" with an overshoot in the region of the CERN SPS data was obtained. 
The "horn" feature (which is also observed for other strange-to-nonstrange hadron ratios: $\Lambda/\pi^+$, $\Xi/\pi^+$, $\Omega^-/\pi^+$) can be understood 
from the fact that the freezeout line in the phase diagram is curved (similar to the pseudocritical line) while the lines of constant $K^+/\pi^+$ are straight and 
just one of them (for the maximum of the $K^+/\pi^+$ ratio) is a tangent to the freezeout line. 
For larger values of $K^+/\pi^+$, there are no intersections, for smaller ones, there are two of them. 
For the HRG model, see Figs. 1 and 4 of Ref.~\cite{Oeschler:2007bj}, where the similarity of the $K^+/\pi^+$ ratio with the Wroblewski factor 
$\lambda_s=2\langle s\bar s\rangle/(\langle u \bar u\rangle + \langle d \bar d\rangle)$ was noticed.
For the PNJL model, this behaviour has been shown in Fig.~7 of \cite{Friesen:2018ojv} in the BW approximation and in Fig.~3 of \cite{Blaschke:2019col}
for the BU approach.
The lines of constant $K^-/\pi^-$ ratio have the opposite slope and cross the freezeout line only once, thus forming a monotonously rising function of the 
increasing freezeout temperature (or $\sqrt{s_{NN}}$).

The comparison of the statistical model result for the energy dependence of the $K^+/\pi^+$ ratio with experiment has been improved by including more higher-lying resonances and the $\sigma$-meson into the HRG model, since the hadron resonance decays populate mainly pion final states so that the ratios of strange particles to pions get somewhat suppressed which remedies the problem with overshooting the data in the SPS energy range \cite{Andronic:2005yp}.
This down-feeding into the pion final state can be seen as a nonequilibrium effect that is not only important to produce the "horn" effect but also the enhancement in the population of low-momentum pion states relative to a thermal equilibrium distribution (without resonance decays) that was observed already in first heavy-ion collision experiments at the CERN-SPS. 
The effect of low-momentum pion enhancement could be successfully described by adopting a pion chemical potential of the order of the pion mass 
$\mu_\pi\approx m_\pi$ \cite{Kataja:1990tp}. 
In the heavy-ion collisions with still higher pion multiplicities at RHIC and LHC, the effect of low-momentum pion enhancement became still more pronounced and 
can no longer be adequately explained with resonance decays. Instead, the nonequilibrium pion distribution with a pion chemical potential became a favorable 
upgrade of the statistical hadronisation model \cite{Begun:2013nga}. 

The proper explanation of the origin of such a chemical potential has been given within the Zubarev approach to nonequilibrium statistical thermodynamics 
that introduces the concept of a nonequilibrium statistical operator \cite{Blaschke:2020afk}.
The number of produced pions is a quasi-conserved number for the duration of the collision event until its detection and has therefore to be included as an additional observable that characterizes the nonequilibrium state into the statistical operator with the pion chemical potential as the Lagrange multiplier 
conjugate to it. 

According to the nonequilibrium statistical models the ratio of the yields of mesons, such as the $K/\pi$ ratios which were obtained in the midrapidity range, 
can be calculated in terms of the ratio of the number densities of mesons ($K^\pm/\pi^\pm = n_{K^\pm}/n_{\pi^{\pm}}$) with
\begin{eqnarray}
\label{eq:nM}
n_{M} &=& d_M \int_0^\infty \frac{\rm d^3 {\bf q}}{(2\pi)^3}g_M(E_M),
\end{eqnarray}
where $g_M(E_M) = (e^{(E_M- \mu_{M})/T}-1)^{-1}$ is the Bose-Einstein distribution function, $E_M=\sqrt{q^2+m_M^2}$ and $\mu_M$ is the chemical potential for the meson species $M$ with the degeneracy factor $d_M$.
The chemical potential for pions as a parameter characterizing the nonequilibrium state has been chosen as a constant close to the pion mass, e.g., 
$\mu_\pi = 0.135$ GeV, following the works \cite{Kataja:1990tp,Begun:2013nga,Naskret:2015pna}. 
A similar nonequilibrium chemical potential can be discussed for kaons too \cite{Kataja:1990tp}, but will not be considered here.
Instead, the chemical potential for kaons is defined here by their chemical composition, i.e. by their quark content, so that 
$\mu_{K^+} =\mu_u-\mu_s$ and $\mu_{K^-} =\mu_s-\mu_u= - \mu_{K^+}$.

For the generalized Beth-Uhlenbeck approach we have the following expression for the meson partial number density as off-shell generalization of the number density of the bosonic species,
\begin{eqnarray}
\label{eq:BU}
n_M(T)&=& 
d_M\int\frac{{ d}^3 \mathbf{q}}{(2\pi)^3} \int \frac{d\omega}{2\pi} g_M(\omega)\frac{d\delta_{M}(\omega,\mathbf{q} )}{d\omega}
\\
&=& \frac{d_{\rm M}}{T} \int\frac{{d}q\, q^2}{2\pi^2}\int_0^\infty\frac{{d}{\omega}}{2\pi} g_M(\omega )(1+g_M(\omega))\delta_M({\omega}),
\label{eq:BUpartial}
\end{eqnarray}
where  $\delta_M(\omega)=\delta_{M}(\omega,\mathbf{0})$ is the meson phase shift that was calculated for mesons at rest, but for which Lorentz-boost invariance $\omega=\sqrt{q^2+{m}^2}$  was assumed, see \cite{Blaschke:2013zaa}.
Examples for the behavior of the phase shifts for $K^+$, $K^-$ and $\pi$ mesons are shown in Fig.~\ref{Fig:ShiftsPiKaMiKaPl} for the situation before (left panel) 
and after (right panel) the Mott dissociation transition of the bound state. 

In the Breit-Wigner approximation, $z_M=E_M -i\Gamma/2$, the phase shift for mesons 
is $\delta_M(\omega) = -\arctan [(\Gamma_M/2)/(\omega-E_M)]$, so that
\begin{equation}
\frac{d\delta_M(\omega)}{d\omega}=\frac{1}{2}\frac{\Gamma_M}{(\omega-E_M)^2+\Gamma_M^2/4}.
\end{equation}
In this case the BU formula for the meson densities (\ref{eq:BU}) reduces to \cite{Andronic:2005yp}
\begin{eqnarray}
\label{eq:n-BW}
n_M(T)&=& 
d_M\int\frac{{ d}^3 \mathbf{q}}{(2\pi)^3} \int_{-\infty}^{\infty} \frac{d\omega}{2\pi} g_M(\omega)\frac{\Gamma_M}{(\omega-E_M)^2+\Gamma_M^2/4}.
\end{eqnarray}
When the mesonic states can be considered on the energy shell (as stable states with infinite lifetime), then for $\Gamma_M\to 0$ Eq.~(\ref{eq:n-BW})
reduces to (\ref{eq:nM}).

\begin{figure}[!th]
\centerline{
\resizebox{0.5\columnwidth}{!}{%
\includegraphics{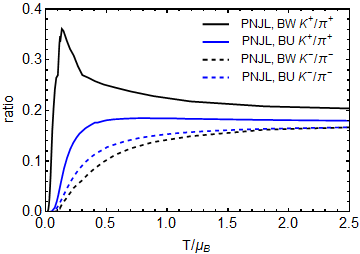}}
\resizebox{0.5\columnwidth}{!}{%
\includegraphics{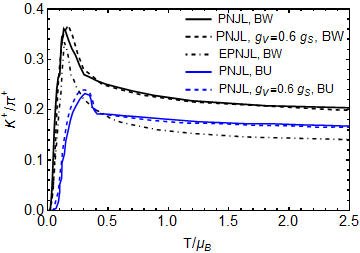}}
}
\caption {Comparison of kaon-to-pion ratios as a function of the specific entropy variable $x=T/\mu_B$ in the BW approximation (black lines) and in the BU approach (blue lines).
Left panel: PNJL model with $\mu_\pi=135$ MeV and $\mu_s=0.55 \mu_u$ shows a horn effect for $K^+/\pi^+$ (solid lines) in the BW approximation but not for the BU approach; $K^-/\pi^-$ (dashed lines) is both approximations a monotonously rising function.
Right panel: for $K^+/\pi^+$ (solid lines) a horn effect is obtained also in the BU approach when $\mu_s$ and $\mu_\pi$ are properly chosen functions of x, see Figs. \ref{nKnpi_ratioWithExp} - \ref{nKnpi_ratioWithExpEnt}. The presence (solid lines) or absence (dashed lines, for $g_V=0.6 g_S$) of a critical endpoint in the phase diagram of the model has a minor influence on the $x$-dependence of  $K^+/\pi^+$. Lowering the pseudocritical chiral restoration temperature towards the chemical freezeout line in the EPNJL model leads to a more rapid drop of the above the horn in the BW approximation, but has no visible effect in the BU approach, see Figs. \ref{nKnpi_ratioWithExp} - \ref{nKnpi_ratioWithExpEnt}. For a detailed discussion see text.}
\label{nKnpi_ratio}
\end{figure}

In the following, we discuss the results for the $K/\pi$ ratios that were obtained by using the above relations for the meson densities in the BU and BW 
approximations along the chiral restoration line that we identify as a proxy for the chemical freezeout.
In order to compare these results with the experimental results we use the variable  $x=T/\mu_B$ instead of $\sqrt{s_{NN}}$, where (T, $\mu_B$) are 
taken along the freezeout line in the phase diagram. 
The rationale for this is the observation that in Lattice QCD \cite{Ejiri:2005uv,Schmidt:1900zza,Ratti:2019tvj} as well as 
within NJL-type models \cite{Motta:2020cbr} the lines of constant entropy per baryon along which the QGP fireball evolves towards hadronization and chemical freezeout are very well approximated by straight lines in the $T-\mu_B$ plane.
Therefore, it is straightforward to use the constant slope $x=T/\mu_B$ of these lines of constant specific entropy as a variable that characterizes the 
initial conditions of the hydrodynamic evolution of the heavy-ion collision and is equivalent to the use of the collision energy $\sqrt{s_{NN}}$. 

In Fig.~\ref{nKnpi_ratio} we compare the kaon-to-pion ratios as a function of the specific entropy variable $x=T/\mu_B$ in the BW approximation (black lines) 
with those in the BU approach (blue lines).
In the left panel for the PNJL model with $\mu_\pi=135$ MeV and $\mu_s=0.55 \mu_u$ the BW approximation shows a horn effect for $K^+/\pi^+$ (solid lines) 
but for the BU approach the horn is absent. 
As it has been discussed above, this shall be a consequence of the different slope of the lines of constant $K^+/\pi^+$ in the phase diagram for the BW and BU approximations, respectively. 
We recall that in the BU approach not only the mesonic bound/resonant states contribute to the partial densities, but also the quark-antiquark scattering in the continuum. 
This contribution is absent in the BW approximation which therefore is not in accordance with the Levinson theorem that requires the scattering phase shift to vanish at high energies.  
The $K^-/\pi^-$ ratio shown by dashed lines is in both approximations a monotonously rising function.

\begin{figure}[!h]
	\centerline{
		\resizebox{0.5\columnwidth}{!}{%
			\includegraphics{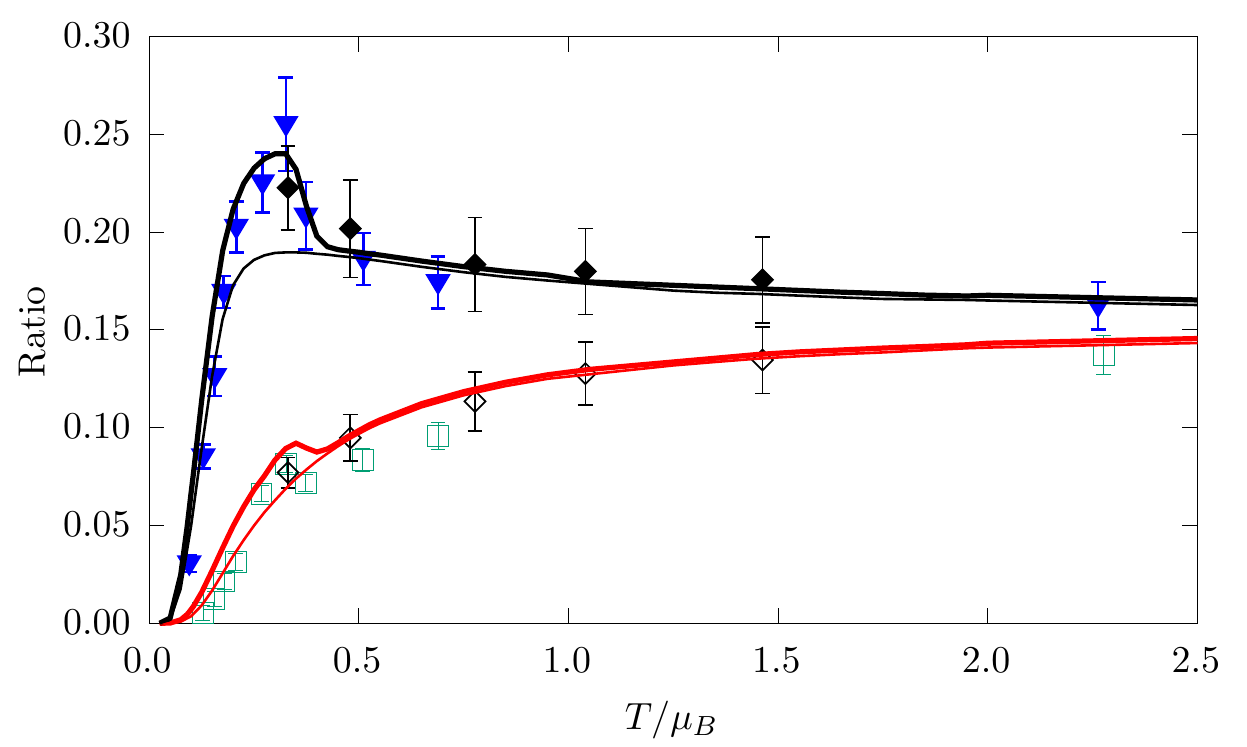}
	}
		\resizebox{0.5\columnwidth}{!}{%
	\includegraphics{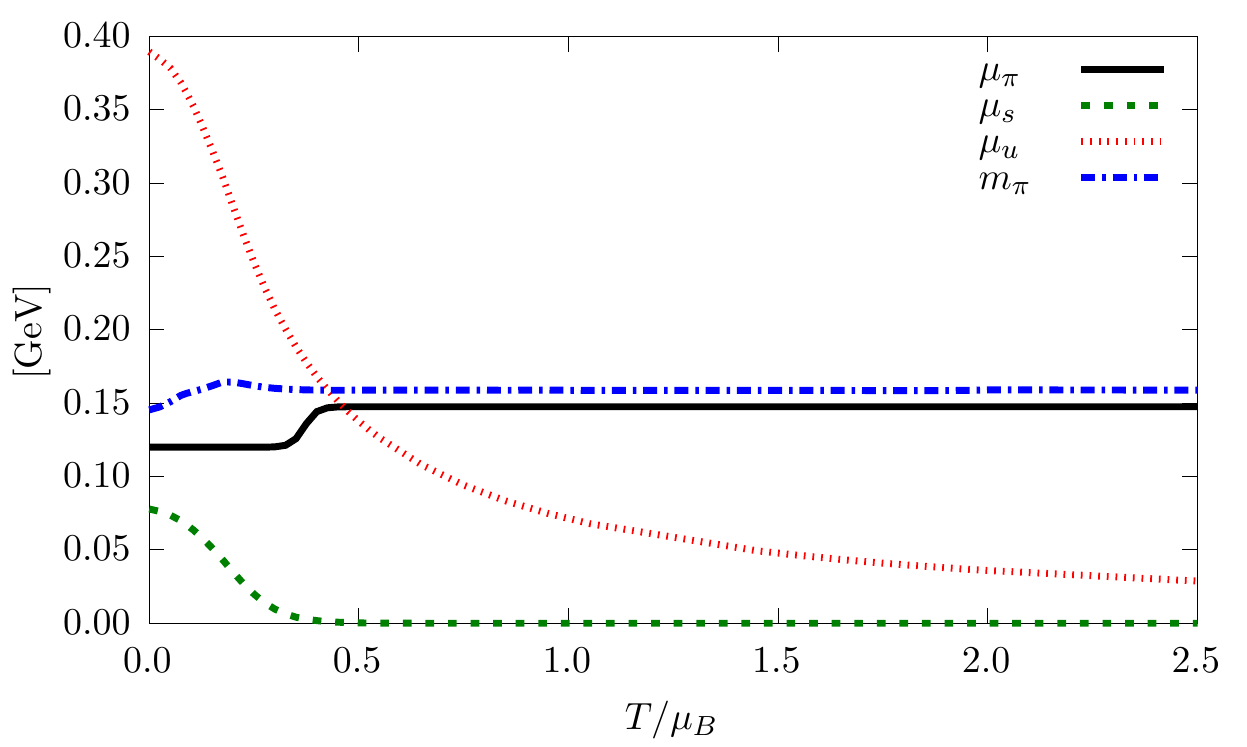}
}}
	\caption {Left panel: The ratios  $K^+/\pi^+$ (black lines) and $K^-/\pi^-$ (red lines) are shown as function of $T/\mu_B$ along the chemical freezeout line 
	for the PNJL+$g_V$ model within the BU approach. Thin lines correspond to the case when $\mu_s=0$ and fixed $\mu_\pi=147.6$ MeV.
	The	thick lines are obtained when $\mu_s$ and $\mu_\pi$ vary with $T/\mu_B$ as shown in right panel.
}
	\label{nKnpi_ratioWithExp}
\end{figure}

\begin{figure}[!h]
	\centerline{
		\resizebox{0.5\columnwidth}{!}{%
			\includegraphics{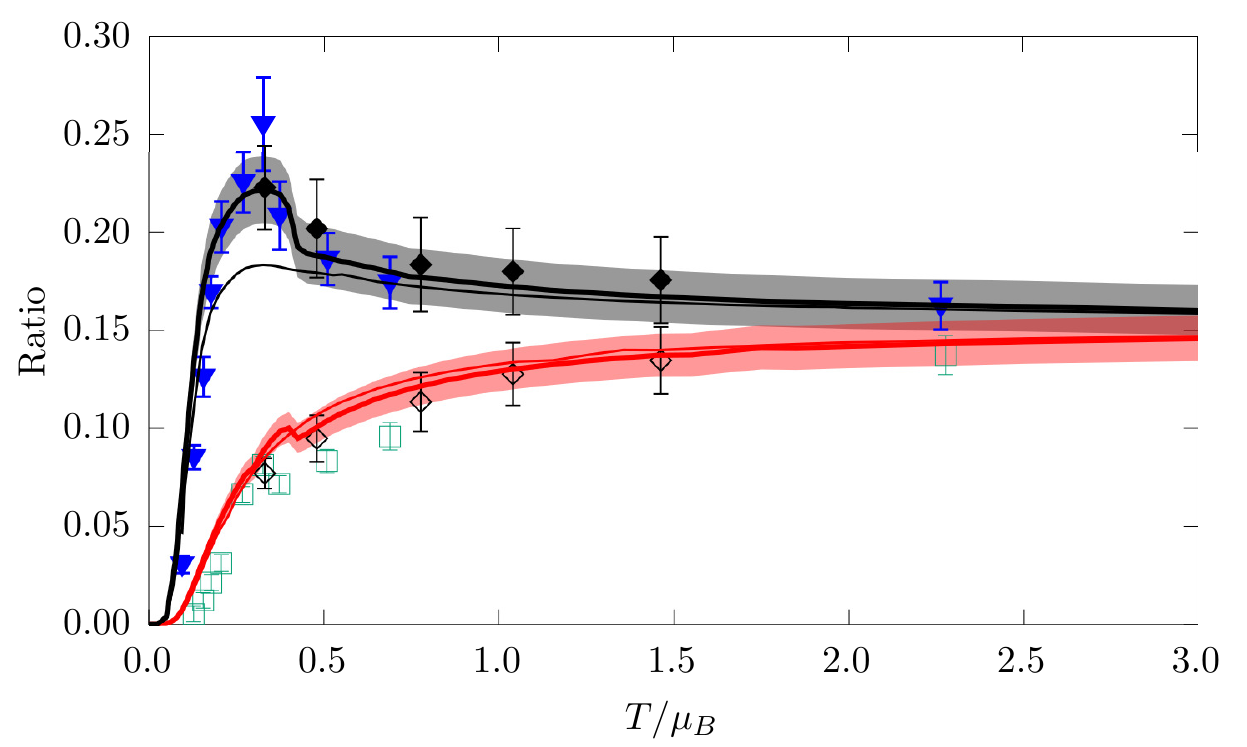}
		}
		\resizebox{0.5\columnwidth}{!}{%
			\includegraphics{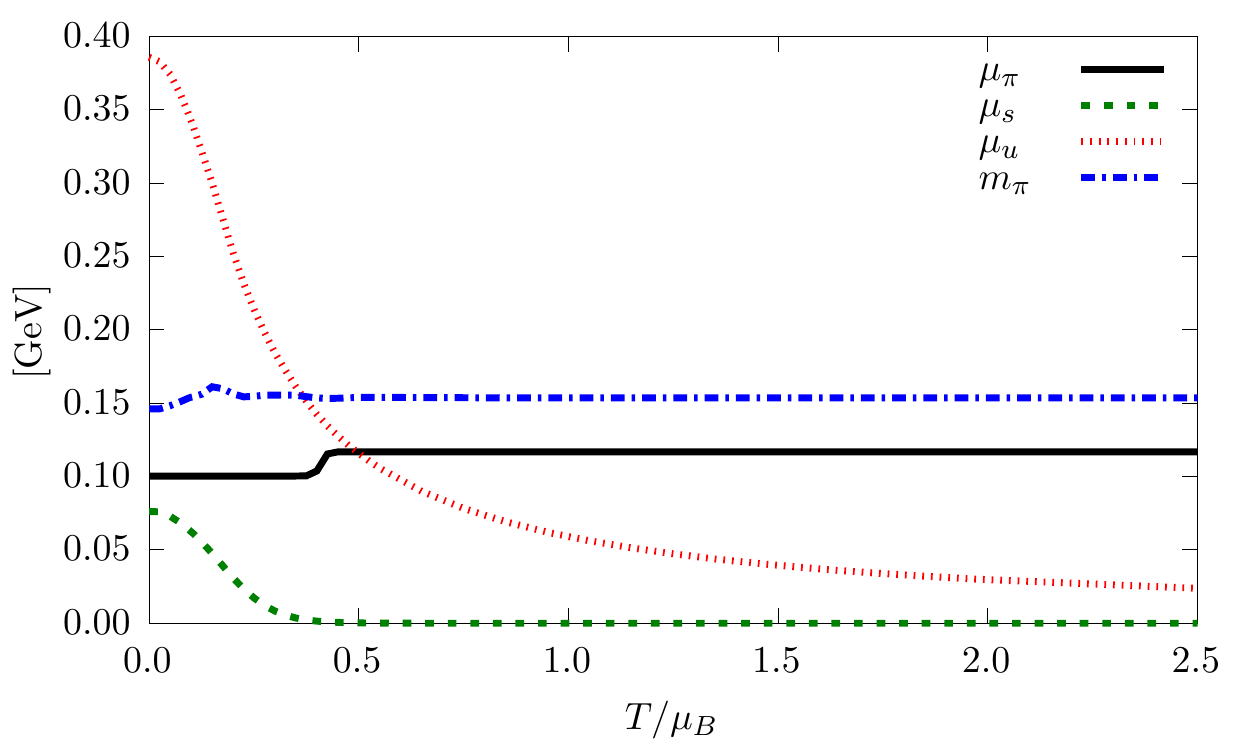}
	}}
	\caption {Same as Fig.~\ref{nKnpi_ratioWithExp} for the  EPNJL+$g_V$ model.
The shaded region corresponds to the variation in fitting $\mu_\pi$ to RHIC and LHC data with their uncertainty band at high $T/\mu_B$. 
	}
	\label{nKnpi_ratioWithExpEntGv}
\end{figure}

\begin{figure}[!h]
	\centerline{
		\resizebox{0.5\columnwidth}{!}{%
			\includegraphics{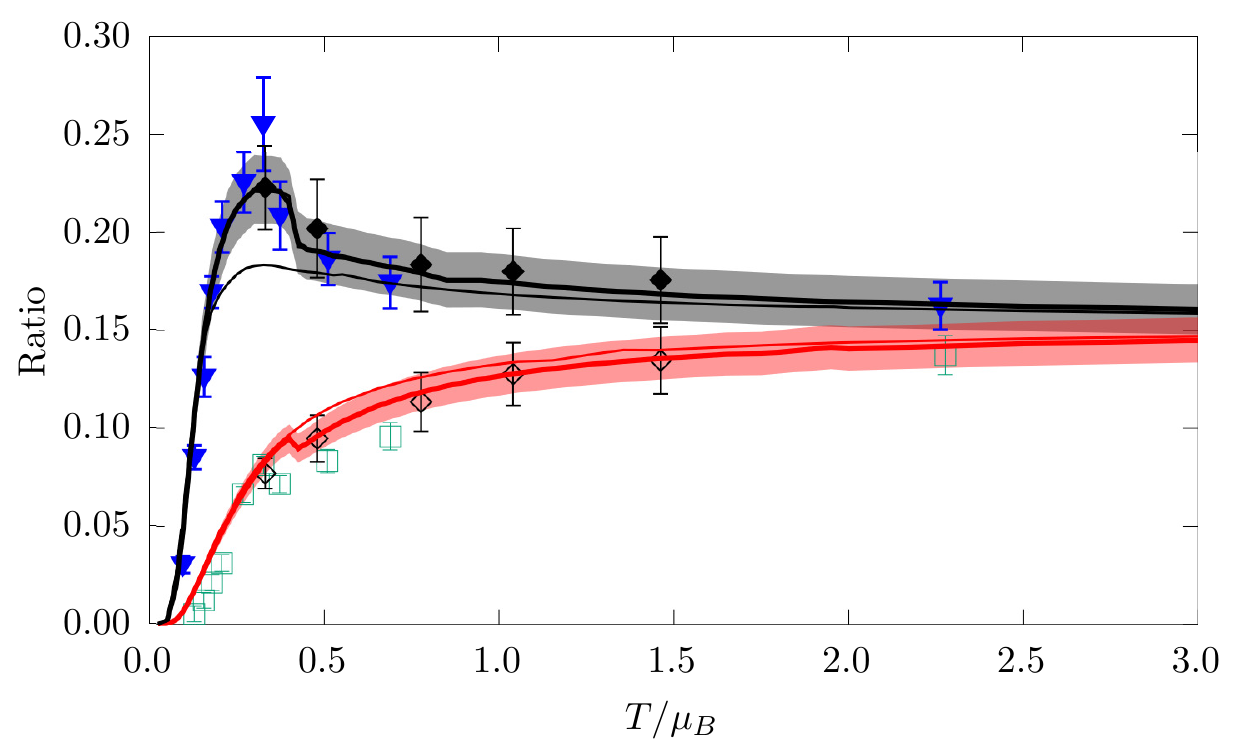}
		}
		\resizebox{0.5\columnwidth}{!}{%
			\includegraphics{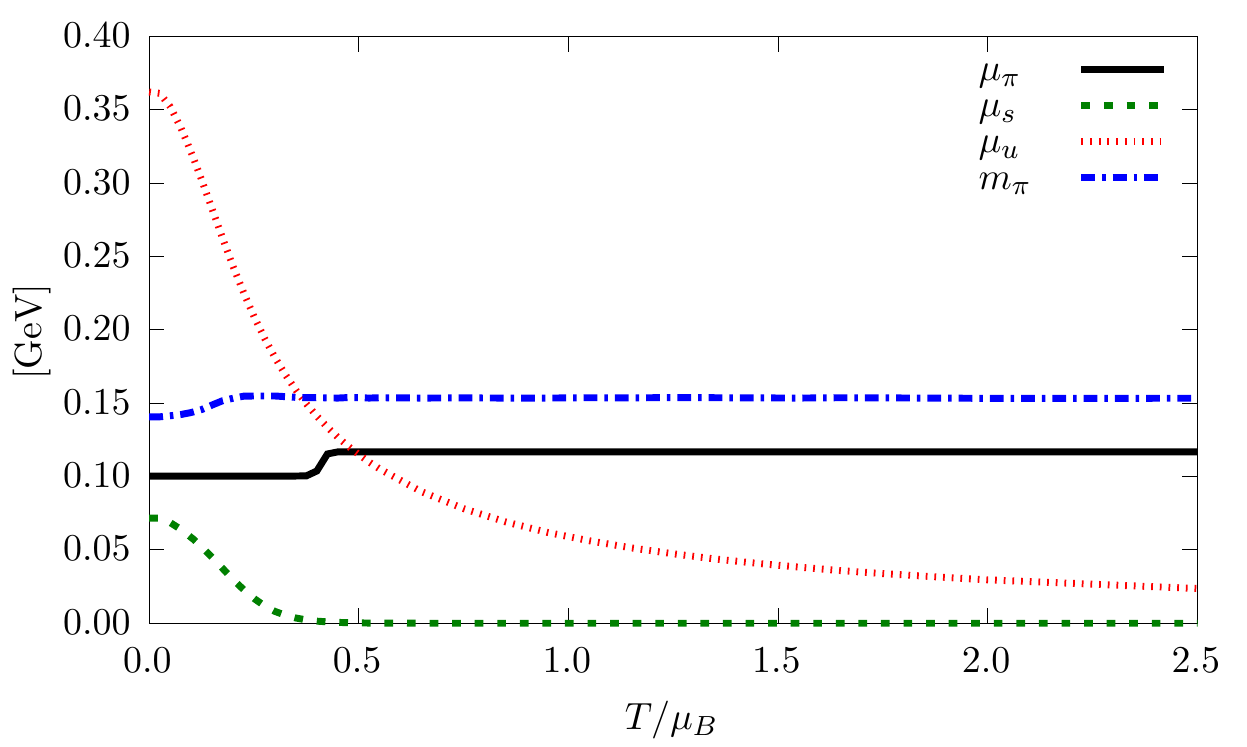}
	}}
	\caption {Same as Fig.~\ref{nKnpi_ratioWithExpEntGv} for the  EPNJL model.
	}
	\label{nKnpi_ratioWithExpEnt}
\end{figure}

In the right panel of Fig.~\ref{nKnpi_ratio} we show the $K^+/\pi^+$ ratio as function of $x$ for the PNJL model without vector coupling (solid lines) which has a critical endpoint in the phase diagram, with vector coupling (dashed lines, for $g_V=0.6 g_S$) when a CEP is absent and for the EPNJL model (dash-dotted line) that has a lower pseudocritcal temperature than the PNJL model, closer to the parametrized chemical freezeout line.
Here a horn effect is obtained also in the BU approach since $\mu_s$ and $\mu_\pi$ are chosen as functions of $x$, see Figs. \ref{nKnpi_ratioWithExp} - \ref{nKnpi_ratioWithExpEnt}. 
The absence (solid lines) or presence (dashed lines) of the vector coupling $g_V=0.6~g_S$, which entails the presence or absence of a critical endpoint in the phase diagram of the model has a minor influence on the $x$-dependence of  $K^+/\pi^+$. 
Lowering the pseudocritical temperature of the chiral restoration towards the parametrized chemical freezeout line by using the EPNJL model  leads to a more rapid drop  above the horn in the BW approximation, but has no visible effect in the BU approach, 
see Figs. \ref{nKnpi_ratioWithExp} - \ref{nKnpi_ratioWithExpEnt}.

In the left panels of Figs. \ref{nKnpi_ratioWithExp} - \ref{nKnpi_ratioWithExpEnt} we show as thick lines the description of the experimental data for the 
kaon-to-pion ratios that is achieved by using the PNJL  (Fig.~\ref{nKnpi_ratioWithExp}) and 
EPNJL (Figs.~\ref{nKnpi_ratioWithExpEntGv}, \ref{nKnpi_ratioWithExpEnt}) models to calculate the densities of pions and kaons the BU approach when 
the nonequilibrium pion chemical potential $\mu_\pi=\mu_\pi(x)$ and strange quark chemical potential $\mu_s=\mu_s(x)$ are assumed to depend on the 
specific entropy variable $x$ as shown in the corresponding right panels. 
For their $x$- dependence we have adopted trial functions of the Woods-Saxon form
\begin{eqnarray}
\label{eq:mu-pi}
\mu_\pi(x) &=& \mu_\pi^{\mathrm{min}}  + \frac{\mu_\pi^{\mathrm{max}} - \mu_\pi^{\mathrm{min}}}{1 + \exp(-(x - x_\pi^{\mathrm{th}})/\Delta x_\pi)) }, 
\\
\mu_s(x) &=&  \frac{\mu_s^{\mathrm{max}}}{1 + \exp(-(x - x_s^{\mathrm{th}})/\Delta x_s)) } .
\label{eq:mu-s}
\end{eqnarray}
The best values of the parameters for the PNJL (EPNJL) model are 
$\mu_\pi^{\mathrm{max}}= 147.6\pm10~(107\pm10)$ MeV, 
$\mu_\pi^{\mathrm{min}}=120~(92)$ MeV,
$x_\pi^{\mathrm{th}}=0.370~(0.409)$, 
$\Delta x_\pi = 0.015~(0.00685)$.
The parameter values
$\mu_s^{\mathrm{max}}/\mu_u^{\mathrm{crit}}= 0.205$, 
$x_s^{\mathrm{th}}=0.223$, $\Delta x_s = 0.06$
hold for both models.
The thin lines on the left panels of Figs. \ref{nKnpi_ratioWithExp} - \ref{nKnpi_ratioWithExpEnt} are obtained for $\mu_\pi=\mu_\pi^{\mathrm{max}}={\rm const}$
and $\mu_s=0$.

The expressions (\ref{eq:mu-pi}) and (\ref{eq:mu-s}) describe the rising pion chemical potential and decreasing strange quark chemical potential, respectively. 
Both of them have  positions and widths of the transition regions which are controlled by parameters. 
For large $x$ the pion chemical potential is $\mu_\pi^{\mathrm{max}}$ while the strange quark chemical potential is zero. 
One can easily estimate the error band for $\mu_\pi^{\mathrm{max}}$ from the variance of the RHIC or LHC data. 
The result for the horn is almost insensitive to the width $\Delta x_\pi$, namely changes of this variable by a factor of 2 will not lead to visible differences.
 The functions $\mu_u(x)$ and  $m_\pi(x)$ along the line of the (pseudo-)critical temperature (which we use as proxy for the chemical freezeout) are results 
 of the calculations within the PNJL and EPNJL models.
 
%
Additionally, the number of negatively charged pions is larger than positive one, so it is possible to naively use the correction 
$n_{\pi^-}/n_{\pi^+}= 1.0+0.0134/[0.023+(x - 0.023)^2]$, fitted to the experimental data \cite{Adamczyk:2017iwn}.
After that one can fit the value of pion chemical potential in order to reproduce the large $x$ points. 

In Fig. \ref{nKnpi_ratioWithExp} we show for the PNJL+$g_V$ model the $K^+/\pi^+$ (black lines) and $K^-/\pi^-$ (red lines) as a function of the specific entropy variable $x=T/\mu_B$ when the BU approach is applied and the pion and strange quark chemical potentials are either constant (thin lines) or given as functions of $x$ (right panel) so that the experimental data can be fitted (thick lines).
Figs.  \ref{nKnpi_ratioWithExpEntGv} and \ref{nKnpi_ratioWithExpEnt} show the same, but for the EPNJL model with and without  vector meson coupling, respectively.
Comparing these three Figures with each other we see no significant differences in the description of the kaon-to-pion ratios despite the fact that the 
pseudocritical (freezeout) temperature for the PNJL model used in Fig.~\ref{nKnpi_ratioWithExp} is up to 40 MeV higher than that for the EPNJL model used in 
Figs.  \ref{nKnpi_ratioWithExpEntGv} and \ref{nKnpi_ratioWithExpEnt}. 
From this example one can conclude that the effect which a shift of the freeze-out line in the phase diagram would have on the $K^+/\pi^+$ ratio can be entirely
compensated by refitting the nonequilibrium pion chemical potential. 
Furthermore, the absence or presence of a critical point (Fig. \ref{nKnpi_ratioWithExpEntGv} vs. Fig. \ref{nKnpi_ratioWithExpEnt}) plays no decisive role 
in the description of the kaon-to pion ratios.
This is understood because the horn originates from the relative positions of the almost straight lines of the kaon-to-pion ratios and their slope to the curved 
freezeout line in the phase diagram while the presence or absence of the critical point has no effect on these features (see the corresponding figures in Refs. \cite{Friesen:2018ojv,Blaschke:2019col,Andronic:2009gj}).  

We note here that the values $\mu_\pi^{\mathrm{max}}= 147.6\pm10~(107\pm10)$ MeV for the nonequilibrium pion chemical potential that are obtained 
from fitting the kaon-to-pion ratio at high energies do bracket the values that were obtained earlier from fits to the pion transverse momentum spectra 
$\mu_\pi = 120~(134.9)$ MeV at SPS (LHC) energies in Ref.~\cite{Kataja:1990tp} (Ref.~\cite{Begun:2013nga}).
The value of the pion chemical potential can be larger than vacuum value of pion mass but shall be below the value of  pion mass on phase transition line that 
is obtained within the PNJL and EPNJL models. 

Since the theoretical concept of introducing a pion chemical potential is the nonequilibrium generalization of the Gibbs ensembles within Zubarev's method 
of the nonequilibrium statistical operator \cite{Blaschke:2020afk}, it is natural to assume a dependence of this Lagrangian multiplier on the pion multiplicities, 
which are directly related collision energy and the specific entropy variable $x$.
Consistent with this new explanation of the horn effect by a step-like increase in the pion chemical potential at $\sqrt{s_{NN}}=8$ GeV is that within the 
thermal statistical model at this very collision energy the entropy density of the system changes its character from being baryon-dominated to meson-dominated
\cite{Cleymans:2004hj}.

 
\section{Conclusions}
\label{sec:concl}

In this work, we  used the SU(3) PNJL and EPNJL models to describe quark matter with pseudoscalar meson excitations at finite temperature and density. 
We demonstrated that there is a significant difference in the description of partial densities of pions and kaons in the framework of these models, when instead
of the standard complex mass pole (Breit-Wigner) approximation for the spectral properties of the mesons the Beth-Uhlenbeck approach is used, that introduces the notion of phase shifts accounting not only for the quark-antiquark bound (resonant) states but also for the scattering states in the continuum.  

We investigated the kaon-to-pion ratios as a function of the variable $x=T/\mu_B$, where T and $\mu_B$ were chosen along 
the line of the (pseudo-)critical temperature as a proxy for the chemical freezeout line in the phase diagram. 
The usage of the variable $x$ is motivated by the observation that along the straight lines of constant $x$ in the phase diagram the value of the specific entropy
$S/N_B=s/n$ is constant and belongs to a conserved quantity in the hydrodynamic evolution of the hadronizing fireball created in central nucleus-nucleus collisions.

In the thermal statistical equilibrium model of hadron production which successfully explains the yields and the ratios of produced hadrons by the two parameters, 
$T$ and $\mu_B$ at chemical freezeout for a nucleus-nucleus collision with a given energy $\sqrt{s_{NN}}$ in the nucleon-nucleon center of mass system, the 
position of the "horn" in the $K^+/\pi^+$ ratio at $\sqrt{s_{NN}}=8$ GeV corresponds to a value $x_{\rm horn}=0.33$. 
While often the horn has been discussed in connection with the position of a critical endpoint in the QCD phase diagram, we point out that if such a correlation occurs in a model description it is accidental. 
According to our analyses in previous works as well as in the present study the key to understanding the horn lies in the position and the slope of the almost straight lines of constant $K^+/\pi^+$ and $K^-/\pi^-$ ratios relative to the curved freezeout line which both are almost unaffected by the appearance of 
a  change from crossover to first order transition. 
According to recent results from lattice QCD the critical endpoint, if it exists, must lie at temperatures below that of the chiral phase transition in the limit of massless quarks which was found to be $T_c^0=132^{+3}_{-6}$ MeV \cite{Ding:2019prx}.
If we adopt the chemical freezeout line as the absolute lower limit of the chiral restoration transition, the corresponding chemical potential 
$\mu_{B,c}=477^{+34}_{-27}$ MeV
at this temperature is a lower limit on the chemical potential of the CEP and an upper limit for the $x$-variable at the CEP (if it exists at all) is obtained as 
$x_{\rm CEP}<0.277^{+0.023}_{-0.033}$, which is still below the value at the horn.  

In the present work we have suggested that the sharpness of the horn effect in the $K^+/\pi^+$ ratio  is well explained by a Bose-enhanced pion production
for $x>x_{\rm horn}$, i.e. for heavy-ion collisions with $\sqrt{s_{NN}}>8$ GeV, in the region of meson dominance, where also a low-momentum enhancement of 
pion production has been observed. 
Such an effect is best described by a nonequilibrium pion distribution function, which according to Zubarev's concept of the nonequilibrium statistical operator 
requires an additional Lagrange multiplier, the pion chemical potential, for a consistent description.
Within the Beth-Uhlenbeck approach we have provided fit functions for the $x$-dependence of the pion and strange quark chemical potentials that lead to a simultaneous description of the $x$-dependence of both kaon-to-pion ratios in accordance with the experimental data, with a strong horn effect for the 
$K^+/\pi^+$ ratio.   

We suggest that in a subsequent work the simultaneous description of the meson momentum spectra should be performed in order to check whether the 
low-momentum pion enhancement due to the fitted $x$- dependence of the nonequilibrium pion chemical potential is in accordance with the experimental data.
The further development of the approach should be devoted to the inclusion of more hadronic channels towards a full hadron resonance gas description, 
including the Mott dissociation of hadrons and the backreaction of these states on the behaviour of the chiral condensate. 
Such a study is beyond the mean field level of description and has the potential to improve the quantitative correspondence of the theoretical model with the 
lattice QCD results on the chiral restoration transition at zero and nonvanishing baryochemical potentials in the QCD phase diagram.

\subsection*{Acknowledgement}
We acknowledge a discussion with F. Karsch and C. Schmidt about the  Lattice QCD results for the entropy per baryon trajectories and critical endpoint position 
in the QCD phase diagram. We thank G. R{\"o}pke for his comments on the quasi-chemical potential for pions in nonequilibrium systems, L. Turko for his remarks about the experimental situation with the "horn" effect and M. Lewicki for his careful reading and discussion of the manuscript. 
This work was supported by the Russian Fund for Basic Research (RFBR) under grant no. 18-02-40137.

\end{document}